\documentstyle[12pt]{article}

\begin{document}
\begin{titlepage}
\setcounter{page}{1}
\title{Vacuum polarization induced by a uniformly accelerated charge}
\author{B. Linet \\
\mbox{\small Laboratoire de Gravitation et Cosmologie Relativistes} \\
\mbox{\small CNRS/URA 769, Universit\'{e} Pierre et Marie Curie} \\
\mbox{\small Tour 22/12, Bo\^{\i}te Courrier 142} \\
\mbox{\small 4, Place Jussieu, 75252 PARIS CEDEX, France}}
\maketitle
\begin{abstract}
We consider a point charge fixed in the Rindler coordinates which describe a
uniformly accelerated frame. We determine an integral expression of the
induced charge density due to the vacuum polarization at the first order in
the fine structure constant. In the case where the acceleration is weak,
we give explicitly the induced electrostatic potential.
\end{abstract}
\end{titlepage}

\section{Introduction}
There has been much interest for a long time in the study of classical and
quantum problems in a uniformly accelerated frame. For example, quantum field
theory in such a frame yields the Unruh effect. The present paper is
concerned with the vacuum polarization due to a charge fixed in a uniformly
accelerated frame. So far as we know, the induced vector potential
has not been determined in this case.

When the pair creation is neglected, the induced current $<j^{\mu}>$
resulting from the vacuum polarization in an external current $j^{\mu}$ has
been determined by Serber (1935) at the first order in the fine structure
constant $\alpha$, making use of the Fourier transform.
Schwinger (1949) has latter given an equivalent integral expression with the
aid of the half-sum of advanced and retarded Green function
${\overline \triangle}(x,x')$. However, the direct application of
these formulas to the case of the current of a uniformly accelerated charge
seems too difficult.

It is of course natural to analyze this problem in a uniformly accelerated
frame described by the Rindler coordinates in which the charge appears as
fixed. In consequence, one should infer that there exists only an
induced charge density in this frame resulting from the vacuum polarization.
Unfortunately, the above mentioned formulas cannot covariantly
write down. However, the Schwinger formula giving the induced current can
be developped in power series in $1/m^{2}$, $m$ being the mass
of the electron ($\hbar =c=1$), and this series
may be rewritten in a manifestly covariant manner. As a consequence of this,
we will derive an integral expression for the induced charge density with
the aid of the Green function for a certain operator expressed in the
Rindler coordinates.

The plan of the work is as follows. In section 2, we recall the basic results
of the vacuum polarization in the first order in $\alpha$. Then we obtain in
section 3 the covariant expression of the induced current in the form of a
power series in $1/m^{2}$. We specialize this result for the case of the
Rindler coordinates in section 4. In section 5, from this we deduce an
integral expression of the induced electrostatic potential for a fixed point
charge. We add in section 6 some concluding remarks.

\section{Schwinger's formula}
In inertial coordinates $(x^{0},x^{i})$ the Minkowskian metric has
the expression
\begin{equation}
\label{1.0}
ds^{2}=-(dx^{0})^{2}+(dx^{1})^{2}+(dx^{2})^{2}+(dx^{3})^{2}
\end{equation}
The Maxwell equations for the vector potential $A_{\mu}$ are
\begin{equation}
\label{1.2a}
\Box A_{\mu}=j_{\mu} \quad {\rm with} \quad \partial_{\mu}A^{\mu}=0
\end{equation}
where $j^{\mu}$ is the current which is conserved. We now introduce
the half-sum of advanced and
retarded Green functions ${\overline \triangle}(x,x')$ for the equation
\begin{equation}
\label{1.1}
(\Box_{x}-m^{2}){\overline \triangle}=-\delta^{(4)}(x,x')
\end{equation}
It has the explicit expression
\begin{eqnarray}
\label{1.1a}
{\overline \triangle}(x,x')= \frac{\delta (\lambda )}{4\pi}-\frac{m^{2}}
{8\pi} \left \{\begin{array}{lll}
J_{1}(m\lambda^{1/2})/m\lambda^{1/2} \quad & & {\rm for} \quad \lambda >0 \\
0 \quad & & {\rm for} \quad \lambda <0
\end{array} \right.
\end{eqnarray}
where $\lambda =(x^{0}-x'^{0})^{2}-(x^{1}-x'^{1})^{2}-(x^{2}-x'^{2})^{2}
-(x^{3}-x'^{3})^{2}$,
$J_{n}$ being the Bessel functions.

Schwinger (1949) has shown that the induced current
$<j^{\mu}>$ due to the vacuum polarization in an external current $j^{\mu}$
can be calculated by an integral containing ${\overline \triangle}(x,x')$;
he has found
\begin{eqnarray}
\label{1.2}
\nonumber <j_{\mu}(x)>&=&-\frac{4\alpha}{\pi}\int dx'^{0}dx'^{1}dx'^{2}
dx'^{3}\int_{0}^{1}dv{\overline \triangle} [\frac{2}
{(1-v^{2})^{1/2}}(x-x')] \\
& &\times \frac{1-v^{2}/3}{(1-v^{2})^{2}}v^{2}\Box_{x'}j_{\mu}(x')
\end{eqnarray}
The induced vector potential $<A_{\mu}>$ is then determined from Maxwell's
equations (\ref{1.2a}) with current (\ref{1.2}); we have immediately
\begin{eqnarray}
\label{1.2c}
\nonumber <A_{\mu}(x)>&=&-\frac{4\alpha}{\pi}\int dx'^{0}dx'^{1}dx'^{2}
dx'^{3} \int_{0}^{1}dv{\overline \triangle}[\frac{2}
{(1-v^{2})^{1/2}}(x-x')] \\
& &\times \frac{1-v^{2}/3}{(1-v^{2})^{2}}v^{2}j_{\mu}(x')
\end{eqnarray}

In the case of a point charge at rest in inertial coordinates,
the non-vanishing component of the external current $j^{\mu}$ is
the charge density $j^{0}$
\begin{equation}
\label{4.1}
j^{0}(x^{i})=e\delta (x^{1})\delta (x^{2})\delta (x^{3})
\end{equation}
where $e$ is the charge. The corresponding electrostatic potential
$A_{0}$ is
\begin{equation}
\label{4.1a}
A_{0}(x^{i})=\frac{e}{4\pi r}
\end{equation}
where $r=\sqrt{(x^{1})^{2}+(x^{2})^{2}+(x^{3})^{2}}$. With a static external
source, one can perform the integration with respect to the variable $x'^{0}$
in Schwinger's formula
(\ref{1.2}). So, the useful quantity is now the Green function for
the operator $\triangle-m^{2}$. Finally formula (\ref{1.2c}), giving the
induced electrostatic potential $<A_{0}>$ for current (\ref{4.1}), reduces to
\begin{equation}
\label{4.4}
<A_{0}(x^{i})>=\frac{e}{4\pi r}\frac{\alpha}{\pi}\int_{0}^{1}dv
\exp [-\frac{2mr}{(1-v^{2})^{1/2}}]
\frac{1-v^{2}/3}{1-v^{2}}v^{2}
\end{equation}
This modification of the Coulomb law has a range $1/2m$. The determination
of $<A_{0}>$ has been done by Uehling (1935) but we give the expression
in closed form found by Pauli and Rose (1936). We set
\begin{equation}
\label{4.4a}
<A_{0}(r)>=\frac{e}{4\pi r}U(mr)
\end{equation}
where the function $U$ can be expressed in terms of elementary functions
\begin{eqnarray}
\label{4.5}
\nonumber U(z)&=&\frac{\alpha}{3\pi}[2(\frac{z^{2}}{3}+1)K_{0}(2z)-
\frac{2z}{3}(2z^{2}+5)K_{1}(2z) \\
& &+z(\frac{4z^{2}}{3}+3)Ki_{1}(2z)]
\end{eqnarray}
$K_{n}$ being the modified Bessel functions of second kind and $Ki_{n}$ the
repeated integrals of $K_{0}$.
{}From expression (\ref{4.5}) we obtain easily the asymptotic form of
$U$ for small values of $z$
\begin{equation}
\label{4.6}
U(z)\sim -\frac{2\alpha}{3\pi}(\gamma +\frac{5}{6}+{\rm ln}z)
\end{equation}
where $\gamma$ is Euler's constant.

In the case of a point charge which is uniformly accelerated with an
acceleration $g$, the external current $j^{\mu}$ has the components
\begin{eqnarray}
\label{4.7}
\nonumber j^{0}(x)&=&e\delta [x^{1}-(1/g^{2}+(x^{0})^{2})^{1/2}]
\delta (x^{2})\delta (x^{3}) \\
\nonumber j^{1}(x)&=&e\frac{x^{0}}{(1/g^{2}+(x^{0})^{2})^{1/2}}
\delta [x^{1}-(1/g^{2}+(x^{0})^{2})^{1/2}]
\delta (x^{2})\delta (x^{3}) \\
j^{2}(x)&=&j^{3}(x)=0
\end{eqnarray}
The application of Schwinger's formula (\ref{1.2}) for current (\ref{4.7})
is possible in principle but the actual calculations are too complicated.
\section{Covariant formula in power series in $1/m^{2}$}

Schwinger's formula (\ref{1.2}) can be developped in power series in
$1/m^{2}$. According to equation (\ref{1.1}), we have the relation
\begin{equation}
\label{3.1}
{\overline \triangle}[\frac{2}{(1-v^{2})^{1/2}}x]=\frac{(1-v^{2})^{2}}
{16m^{2}} \delta^{(4)}(x)+\frac{(1-v^{2})}{4m^{2}}
\Box {\overline \triangle}[\frac{2}{(1-v^{2})^{1/2}}x]
\end{equation}
By inserting (\ref{3.1}) into (\ref{1.2}) and by performing an integration
by part, we obtain
\begin{eqnarray}
\label{3.2}
\nonumber <j_{\mu}(x)>&=&-\frac{\alpha}{4\pi m^{2}}\int_{0}^{1}
dv(1-v^{2}/3)v^{2} \ \Box j_{\mu}(x) \\
\nonumber & &-\frac{\alpha}{\pi m^{2}}\int dx'^{0}dx'^{1}dx'^{2}
dx'^{3}\int_{0}^{1}dv
{\overline \triangle}[\frac{2}{(1-v^{2})^{1/2}}(x-x')] \\
& &\times \frac{1-v^{2}/3}{1-v^{2}}v^{2}\Box^{2}_{x'}j_{\mu}(x')
\end{eqnarray}
The term proportional to $\Box j_{\mu}$ in (\ref{3.2}) is the first term of
the power series in $1/m^{2}$.
By inserting again relation (\ref{3.1}) into the second term in (\ref{3.2}),
we will obtain the second term of the power series in $1/m^{2}$, and so on.
Hence the current of polarization has the expression
\begin{equation}
\label{3.3}
<j_{\mu}(x)>=\sum_{n=1}^{\infty}a_{n}\frac{1}{m^{2n}}\Box^{n}j_{\mu}(x)
\end{equation}
where all the coefficients $a_{n}$ can be calculated. In particular we have
\begin{equation}
\label{3.4}
a_{1}=-\frac{\alpha}{15\pi m^{2}}
\end{equation}

Each term of series (\ref{3.3}) is a 1-form that we may covariantly write
down. We now consider an arbitrary coordinate system $(x^{\mu '})$ of the
Minkowski spacetime. The components of the Minkowskian metric are denoted
$g_{\rho '\sigma '}$ and the covariant derivative $\nabla_{\rho '}$.
According to (\ref{3.3}), the induced current $<j_{\mu '}>$
can be expressed in function of the external current $j_{\mu '}$ by the
following power series in $1/m^{2}$
\begin{equation}
\label{3.5}
<j_{\mu '}(x')>=\sum_{n=1}^{\infty}a_{n}\frac{1}{m^{2n}}(g^{\rho '\sigma '}
\nabla_{\rho '}\nabla_{\sigma '}j_{\mu '})^{n}
\end{equation}
where the operators in (\ref{3.5}) are defined by the law of recurrence
\begin{equation}
\label{3.6}
(g^{\rho '\sigma '}\nabla_{\rho '}\nabla_{\sigma '}j_{\mu '})^{n}=
g^{\rho '\sigma '}\nabla_{\rho '}\nabla_{\sigma '}
[(g^{\rho '\sigma '}\nabla_{\rho '}\nabla_{\sigma '}j_{\mu '})^{n-1}]
\quad (n\geq 1)
\end{equation}
To establish this result we have taken into account that the Ricci tensor
vanishes.

The induced vector potential $<A_{\mu '}>$ satisfies the covariant Maxwell
equations
\begin{equation}
\label{3.7}
g^{\rho '\sigma '}\nabla_{\rho '}\nabla_{\sigma '}A_{\mu '}=j_{\mu '}
\quad {\rm with} \quad \nabla_{\mu '}A^{\mu '}=0
\end{equation}
\section{Case of the Rindler coordinates}

The application of Schwinger's formula (\ref{1.2}) to the case of a uniformly
accelerated charge, described by current (\ref{4.7}), seems very difficult
because the problem is time-dependent. But,
we know that the Rindler coordinates $(\xi^{0},\xi^{1} ,\xi^{2},\xi^{3})$,
with $\xi^{1}>0$, describe a uniformly accelerated frame in the Minkowski
spacetime. For an acceleration $g$, the coordinate transform from inertial
coordinates is
\begin{eqnarray}
\label{1.4}
\nonumber x^{0}&=&\xi^{1} \sinh (g\xi^{0}) \\
\nonumber x^{1}&=&\xi^{1} \cosh (g\xi^{0})  \\
x^{2}&=&\xi^{2} \quad {\rm and} \quad x^{3}=\xi^{3}
\end{eqnarray}
In this coordinate system, Minkowskian metric (\ref{1.0}) takes the form
\begin{equation}
\label{1.5}
ds^{2}=-g^{2}(\xi^{1})^{2}(d\xi^{0})^{2}+(d\xi^{1})^{2}+(d\xi^{2})^{2}+
(d\xi^{3})^{2}
\end{equation}
The charge having an acceleration $g$ will be now located at the point
\begin{equation}
\label{1.6}
\xi^{1} =\frac{1}{g} \quad \xi^{2}=\xi^{3}=0
\end{equation}
and its current (\ref{4.7}) has the following components
\begin{eqnarray}
\label{1.7}
\nonumber j^{\xi^{0}}&=&e\delta (\xi^{1} -\frac{1}{g})
\delta (\xi^{2})\delta (\xi^{3}) \\
j^{\xi^{1}}&=&0 \quad {\rm and} \quad j^{\xi^{2}}=j^{\xi^{3}}=0
\end{eqnarray}
in the Rindler coordinates. Consequently, the uniformly accelerated chage
is described by a point charge at rest.

Maxwell's equations (\ref{3.7}) written in the Rindler coordinates
for a static charge density $j^{\xi^{0}}$ reduce to
an equation for the electrostatic potential $A_{\xi^{0}}$
\begin{equation}
\label{2.3}
(\triangle_{\xi}-\frac{1}{\xi^{1}}\frac{\partial}{\partial \xi^{1}})
A_{\xi^{0}}=j_{\xi^{0}}
\end{equation}
where $j_{\xi^{0}}=-(g\xi^{1})^{2} j^{\xi^{0}}$. For a point charge at rest,
$j^{\xi^{0}}$ being given by (\ref{1.7}), the
electrostatic potential $V_{W}$ has been found by Whittaker (1927),
in a slightly different coordinate system, which corresponds to the retarded
solution to the Maxwell equations with current (\ref{4.7}).

In Rindler coordinates, we remark that the operator
$g^{\rho \sigma}\nabla_{\rho}\nabla_{\sigma}j_{\mu}$
applied to a static charge density $j^{\xi^{0}}$ is simple since we have
\begin{eqnarray}
\label{2.4}
\nonumber (g^{\rho \sigma}\nabla_{\rho}\nabla_{\sigma}j_{\xi^{0}})&=&
(\triangle_{\xi}-\frac{1}{\xi^{1}}\frac{\partial}{\partial \xi^{1}})
j_{\xi^{0}} \\
(g^{\rho \sigma}\nabla_{\rho}\nabla_{\sigma}j_{\xi^{1}})&=&(g^{\rho \sigma}
\nabla_{\rho}
\nabla_{\sigma}j_{\xi^{2}})=(g^{\rho \sigma}\nabla_{\rho}
\nabla_{\sigma}j_{\xi^{3}})=0
\end{eqnarray}
As a consequence of properties (\ref{2.4}), expression (\ref{3.5})
of the induced current yields only
a charge density $<j^{\xi^{0}}>$. This fact is natural since we have now a
problem which does not depend on the time.
By defining the operator
\begin{equation}
\label{2.5}
{\cal D}_{\xi}=\triangle_{\xi}-\frac{1}{\xi^{1}}\frac{\partial}
{\partial \xi^{1}}
\end{equation}
we can rewrite the induced charge density as a power series in $1/m^{2}$
\begin{equation}
\label{2.6}
<j_{\xi_{0}}>=\sum_{n=1}^{\infty}a_{n}\frac{1}{m^{2n}}{\cal D}^{n}_{\xi}
j_{\xi^{0}}
\end{equation}
where $j^{\xi^{0}}$ is the external charge density. Maxwell's equations
(\ref{2.3}) are also rewrite in the form
\begin{equation}
\label{2.7}
{\cal D}_{\xi}<A_{\xi^{0}}>=<j_{\xi^{0}}>
\end{equation}

In the Rindler coordinates, the first correction to the Whittaker potential
which is necessary to evaluate the Lamb shift is
\begin{equation}
\label{2.9}
<A_{\xi^{0}}(\xi^{i})>=e\frac{\alpha}{15\pi m^{2}}\delta (\xi^{1}-\frac{1}
{g})\delta (\xi^{2}) \delta (\xi^{3})
\end{equation}
since $a_{1}$ has value (\ref{3.4}).
\section{Vacuum polarization for a charge fixed in the Rindler coordinates}

We now define the Green function ${\cal G}(\xi^{i},\xi '^{i})$ for the
equation
\begin{equation}
\label{5.1}
({\cal D}_{\xi}-m^{2}){\cal G}=-g\xi^{1}\delta^{(3)}(\xi^{i}-\xi '^{i})
\end{equation}
assuming that ${\cal G}(\xi^{i},\xi '^{i})$ vanishes when the points
$\xi^{i}$ and $\xi '^{i}$ are infinitely separated.
Now the operator $1/\xi {\cal D}_{\xi}$ is self-adjoint, therefore the Green
function is symmetric and satisfies the identities
\begin{eqnarray}
\label{5.2}
\nonumber {\cal D}_{\xi}{\cal G}(\xi^{i},\xi '^{i})&=&{\cal D}_{\xi '}
{\cal G}(\xi^{i},\xi '^{i}) \\
\int d\xi^{1}d\xi^{2}d\xi^{3}f(\xi^{i})\frac{1}{\xi^{1}}
{\cal D}_{\xi}g(\xi^{i})&=&
\int d\xi^{1}d\xi^{2}d\xi^{3}g(\xi^{i})\frac{1}{\xi^{1}}
{\cal D}_{\xi}f(\xi^{i})
\end{eqnarray}
where $f$ anf $g$ are two arbitrary functions.

For a static charge density $j^{\xi^{0}}$ we are now in a position to set the
formula giving the induced charge density $<j^{\xi^{0}}>$ under an integral
form
\begin{eqnarray}
\label{5.3}
\nonumber <j_{\xi^{0}}(\xi^{i})>&=&-\frac{\alpha}{\pi}\int
d\xi '^{1}d\xi '^{2}d\xi '^{3}
\int_{0}^{1}dv{\cal G}[\frac{2}{(1-v^{2})^{1/2}}\xi^{i},
\frac{2}{(1-v^{2})^{1/2}}\xi '^{i}] \\
& & \times \frac{1-v^{2}/3}{1-v^{2}}v^{2}\frac{1}{g\xi '^{1}}
{\cal D}_{\xi '}j_{\xi^{0}}(\xi ')
\end{eqnarray}
In order to prove this, we develop formula (\ref{5.3}) in power series in
$1/m^{2}$. We proceed as in section 3. From (\ref{5.1}) we have the relation
\begin{eqnarray}
\label{5.4}
{\cal G}[\frac{2}{(1-v^{2})^{1/2}}\xi^{i} ,\frac{2}{(1-v^{2})^{1/2}}
\xi '^{i}]&=&\frac{1-v^{2}}{4m^{2}}g\xi^{1}
\delta^{(3)}(\xi^{i}-\xi '^{i})+ \\
\nonumber & &\frac{1-v^{2}}{4m^{2}}{\cal D}_{\xi}
{\cal G}[\frac{2}{(1-v^{2})^{1/2}}\xi^{i},
\frac{2}{(1-v^{2})^{1/2}}\xi '^{i}]
\end{eqnarray}
taking into account the specific property of the operator ${\cal D}_{\xi}$.
By inserting (\ref{5.4}) into (\ref{5.3}) and making use identities
(\ref{5.2}), we find
\begin{eqnarray}
\label{5.5}
\nonumber <j_{\xi^{0}}(\xi^{i})>&=&-\frac{\alpha}{4\pi m^{2}}\int_{0}^{1}dv
(1-v^{2}/3)v^{2} \ {\cal D}_{\xi}j_{\xi^{0}}(\xi^{i}) \\
\nonumber & &-\frac{\alpha}{4\pi m^{2}}\int d\xi '^{1}d\xi '^{2}
d\xi '^{3}\int_{0}^{1}dv
{\cal G}[\frac{2}{(1-v^{2})^{1/2}}\xi^{i},\frac{2}{(1-v^{2})^{1/2}}
\xi '^{i}] \\
& &\times (1-v^{2}/3)v^{2}\frac{1}{g\xi '^{1}}
{\cal D}_{\xi '}^{2}j_{\xi^{0}}(\xi '^{i})
\end{eqnarray}
We recognize that the coefficient in front of ${\cal D}_{\xi}j_{\xi^{0}}$ is
$a_{1}$. The repeated applications of relation (\ref{5.4}) yield
power series (\ref{2.6}).
Therefore, we conclude that formula (\ref{5.3}) gives the induced charge
density due to the vacuum polarization. However, we are not worried about the
boundary conditions of the Green function ${\cal G}(\xi^{i},\xi '^{i})$ at
the hypersurface $\xi^{1}=0$ in metric (\ref{1.5}).

We obtain the induced electrostatic potential from
Maxwell's equations (\ref{2.7}) by using again (\ref{5.2})
\begin{eqnarray}
\label{5.6}
\nonumber <A_{\xi^{0}}(\xi^{i})>&=&-\frac{\alpha}{\pi}\int d\xi '^{1}
d\xi '^{2}d\xi '^{3}\int_{0}^{1}dv
{\cal G}[\frac{2}{(1-v^{2})^{1/2}}\xi^{i},\frac{2}{(1-v^{2})^{1/2}}
\xi '^{i}] \\
& & \times \frac{1-v^{2}/3}{1-v^{2}}v^{2}\frac{1}{g\xi '^{1}}
j_{\xi^{0}}(\xi '^{i})
\end{eqnarray}
In the case of charge density (\ref{1.7}) integral (\ref{5.6}) reduces to
\begin{equation}
\label{5.6a}
<A_{\xi^{0}}(\xi^{i})>=\frac{e\alpha}{\pi}\int_{0}^{1}dv
{\cal G}[\frac{2}{(1-v^{2})^{1/2}}\xi^{i},
\frac{2}{(1-v^{2})^{1/2}}\xi_{g}^{i}]\frac{1-v^{2}/3}{1-v^{2}}v^{2}
\end{equation}
where $\xi_{g}^{i}=(1/g,0,0)$. By the inverse transform of coordinates
(\ref{1.4}) we can express the induced vector potential $<A_{\mu}>$ in
inertial coordinates.

The problem is now to determine the Green function
${\cal G}(\xi^{i} ,\xi '^{i})$.
We restrict ourselves to find an expression at the first order in $g$.
To do this, we introduce the new coordinates
\begin{equation}
\label{5.7}
y^{1}=\xi^{1}-\frac{1}{g} \quad , \quad y^{2}=\xi^{2} \quad
{\rm and} \quad y^{3}=\xi^{3}
\end{equation}
With variables (\ref{5.7}), equation (\ref{5.1}) takes the form
\begin{equation}
\label{5.8}
(\triangle_{y}-\frac{g}{1+gy^{1}}\frac{\partial}{\partial y^{1}}-m^{2})
{\cal G}=-(1+gy^{1})\delta^{(3)}(y^{i}-y'^{i})
\end{equation}
and, by keeping the terms linear in $g$, it becomes
\begin{equation}
\label{5.9}
(\triangle_{y}-g\frac{\partial}{\partial y^{1}}-m^{2}){\cal G}=
-(1+gy^{1})\delta^{(3)}(y^{i}-y'^{i})
\end{equation}
The domain of validity of equation (\ref{5.9}) is restricted to $gy^{1}<<1$.
We choose the solution of this equation which reduces to the Green function
for the operator $\triangle -m^{2}$ in the limit where $g$ tends to 0.
We do not touch upon the problem of the global definition of the Green
function ${\cal G}(\xi^{i},\xi '^{i})$. We find
\begin{equation}
\label{5.10}
{\cal G}(y^{i},y'^{i})=\frac{\exp -m\mid y^{i}-y'^{i}\mid}
{4\pi \mid y^{i}-y'^{i}\mid }
(1+\frac{1}{2}gy^{1}+\frac{1}{2}gy'^{1})+O(g^{2})
\end{equation}

We are now in a position to calculate formula (\ref{5.6a}) at the first order
in $g$. We have to perform
$\xi \leadsto 2/(1-v^{2})^{1/2}\xi$ that we write in the coordinates $y^{i}$
\[
1+\frac{1}{2}gy^{1}+\frac{1}{2}gy'^{1} \leadsto \frac{2+gy^{1}+
gy'^{1}}{(1-v^{2})^{1/2}}
\]
So, we obtain
\begin{equation}
\label{5.11a}
<A_{\xi^{0}}(\xi^{i})>\approx \frac{e}{4\pi \epsilon}\frac{\alpha}{\pi}
\int_{0}^{1}dv
\exp [-\frac{2m\epsilon}{(1-v^{2})^{1/2}}]
\frac{1-v^{2}/3}{1-v^{2}}v^{2}(1+\frac{1}{2}gy^{1})
\end{equation}
where $\epsilon =\sqrt{(y^{1})^{2}+(y^{2})^{2}+(y^{3})^{2}}$. According to
expression (\ref{4.4}) of the Uehling potential $<A_{0}>$, we can write
\begin{equation}
\label{5.12}
<A_{\xi^{0}}(y^{i})\approx <A_{0}(y^{i})>+\frac{1}{2}g<A_{0}(y^{i})>y^{1}
\end{equation}

At the first order in $g$, metric (\ref{1.5}) takes the form
\begin{equation}
\label{5.13}
ds^{2}\approx -(1+2gy^{1})(dy^{0})^{2}+(dy^{1})^{2}+(dy^{2})^{2}+(dy^{3})^{2}
\end{equation}
which is valid for $gy^{1}<<1$.
The Whittaker potential is then approximed by
\begin{equation}
\label{5.13a}
V_{W}(y^{i})\approx \frac{e}{4\pi \epsilon}(1+\frac{1}{2}gy^{1})
\end{equation}
The total electrostatic potential $V$, sum of (\ref{5.12}) and (\ref{5.13a}),
generated by a point charge located at $y^{i}=0$ in metric (\ref{5.13}),
taking into account the vacuum polarization at the first order in $\alpha$,
has the expression
\begin{equation}
\label{5.14}
V(y^{i})\approx \frac{e}{4\pi \epsilon}(1+U(m\epsilon)+
\frac{1}{2}gy^{1}+\frac{1}{2}gU(m\epsilon)y^{1})
\end{equation}
where $U$ is given by (\ref{4.5}).
\section{Conclusion}

We have given an integral expression (\ref{5.6a}) for the induced
electrostatic potential with the aid of the Green function for
operator (\ref{5.1}) in the Rindler coordinates. This determination is
obtained from the Schwinger formula in inertial coordinates.  However,
there would be a conceptual interest to derive directly this formula within
the framework of the quantum electrodynamics in Rindler spacetime in order
to discuss the effect of the horizon.

\newpage
{\bf References}

Serber, R. (1935). {\em Physical  Review}, {\bf 48}, 49.

Schwinger, J. (1949). {\em Physical  Review}, {\bf 75}, 651.

Uehling, E.A. (1935). {\em Physical Review}, {\bf 48}, 55.

Pauli, W. and Rose, M.E. (1936). {\em Physical Review,} {\bf 49}, 749.

Whittaker, E.T. (1927). {\em Proceedings of the Royal Society of London
\newline Series A}, {\bf 116}, 726.

\end{document}